\documentclass[]{emulateapj}

\usepackage{natbib,amsmath}

\newcommand{\papertwo}{Updike et al.\, in prep.}
\newcommand{\mnht}{N_{\rm H_2}}
\newcommand{\nht}{$N_{\rm H_2}$}
\newcommand{\texh}{T_{\rm ex}^{\rm H_2}}
\newcommand{\texco}{T_{\rm ex}^{\rm CO}}

\newcommand{\bouv}{\beta_{\rm O}}

\newcommand{\lya}{Ly$\alpha$}

\def\h2{H$_2$}
\def\f0{$F_0$}

\newcommand{\cm}[1]{\, {\rm cm^{#1}}}
\newcommand{\mkms}{{\rm \; km\;s^{-1}}}

\newcommand{\N}[1]{{N({\rm #1})}}

\newcommand{\mnhi}{N_{\rm H~I}}

\newcommand{\nhi}{$N_{\rm HI}$}

\def\nhi{$N_{\rm HI}$}

\begin{document}

\submitted{Accepted to ApJL, December 3 2008}
\title{
The First Positive Detection of Molecular Gas in a GRB Host Galaxy
}
%\title{NV Gas in GRB Afterglow Spectra:
%A Snapshot of Gas Near the Progenitor}
%\author{Blah}
\author{J. X. Prochaska\altaffilmark{1,2},
        Y. Sheffer\altaffilmark{3},
	D.A. Perley\altaffilmark{4},
	J. S. Bloom\altaffilmark{4,5}, 
	L. A. Lopez\altaffilmark{2},
	M. Dessauges-Zavadsky\altaffilmark{6},
	H.-W. Chen\altaffilmark{7},
	A. V. Filippenko\altaffilmark{4}, 
	M. Ganeshalingam\altaffilmark{4}, 
	W. Li\altaffilmark{4}, 
	A. A. Miller\altaffilmark{4}, and
	D. Starr\altaffilmark{4,8}
}

\altaffiltext{1}{University of California Observatories/Lick 
Observatory, University of California, Santa Cruz, CA 95064.} 
\altaffiltext{2}{Department of Astronomy and Astrophysics,
University of California, Santa Cruz, CA 95064.} 
\altaffiltext{3}{Department of Physics and Astronomy,
University of Toledo, Toledo, OH 43606.}
\altaffiltext{4}{Department of Astronomy,
        University of California, Berkeley, CA 94720-3411.}
\altaffiltext{5}{Sloan Research Fellow.}
\altaffiltext{6}{Observatoire de Gen\`eve, 51 Ch. des Maillettes, 
1290 Sauverny, Switzerland.}
\altaffiltext{7}{Department of Astronomy \& Astrophysics, and
Kavli Institute for Cosmological Physics,
5640 S.\ Ellis Ave, Chicago, IL, 60637.}
\altaffiltext{8}{Las Cumbres Observatory, 6740 Cortona Dr. Suite 102,
Santa Barbara, CA 93117.}

%\altaffiltext{3}{Observatoire de Gen\`eve, 51 Ch. des Maillettes, 
%1290 Sauverny, Switzerland.}
%\altaffiltext{4}{Department of Astronomy \& Astrophysics and
%Kavli Institute for Cosmological Physics,
%5640 S.\ Ellis Ave, Chicago, IL, 60637.}

\begin{abstract}
%While radio telescopes routinely detect millimeter-wavelength
%emission from molecular gas
%at redshift $z > 1$, these observations provide limited constraints on the
%properties of the gas.  Here, 
We report on strong H$_2$ and CO
absorption from gas within the
host galaxy of gamma-ray burst (GRB) 080607. 
Analysis of our Keck/LRIS afterglow spectrum reveals a very large
\ion{H}{1} column density ($\mnhi = 10^{22.70 \pm 0.15} \cm{-2}$) and
strong metal-line absorption at $z_{\rm GRB} = 3.0363$
with a roughly solar metallicity. 
We detect a series of $A-X$ bandheads from CO 
and estimate $\N{\rm CO} = 10^{16.5 \pm 0.3} \cm{-2}$ 
and $T_{ex}^{\rm CO} > 100$~K. We
argue that the high excitation temperature results
from UV pumping of the CO gas by the GRB afterglow.
Similarly, we observe H$_2$ absorption via the Lyman-Werner bands
and estimate $\mnht = 10^{21.2 \pm 0.2} \cm{-2}$ with 
$T_{ex}^{\rm H_2} = 10$--300~K.
The afterglow photometry suggests
an extinction law with
$R_V \approx 4$ and $A_V \approx 3.2$ mag and requires the presence of a
modest 2175~\AA\ bump. 
Additionally, modeling of the {\it Swift}/XRT
X-ray spectrum confirms a large column density with 
$N_{\rm H} = 10^{22.58 \pm 0.04} \cm{-2}$.
Remarkably, this molecular gas has
extinction properties, metallicity, and a CO/H$_2$ ratio comparable
to those of translucent molecular clouds of the Milky Way, suggesting
that star formation at high $z$ proceeds 
in similar environments as today. However,
the integrated dust-to-metals ratio is sub-Galactic, suggesting
the dust is primarily associated with the molecular phase while the atomic
gas has a much lower dust-to-gas ratio.  Sightlines like
GRB~080607 serve as powerful probes of nucleosynthesis and
star-forming regions in the young universe and
contribute to the population of ``dark'' GRB afterglows.

\end{abstract} 


\section{Introduction}

Most long-duration, soft-spectrum gamma-ray bursts (GRBs) are thought to 
arise in the death of massive stars 
\citep[see][for a review]{wb06}. 
The evidence for this is both direct, such as the observation of 
a contemporaneous supernova associated with some events 
\citep[e.g.,][]{smg+03,hsm+03}, 
and circumstantial, based on physical associations with star forming 
galaxies \citep[see][]{fls+06}. 
Still, some telltale signatures of massive-star progenitors are rarely, 
if ever, observed. The absence of wind-blown density profiles in the 
circumburst environment \citep{chevalier04} is thought to be due to the 
pre-burst interactions of the wind with the interstellar medium 
\citep[ISM;][]{rgs+05}. 
The absence of Wolf-Rayet signatures in afterglow spectra \citep{cpr+07}
is explained by the  influence of the afterglow light on gas and dust 
within $\sim$100 pc of the GRB event.

Outside the overwhelmingly destructive influence within the inner 100\,pc, 
there are generic signatures of the ISM: \ion{H}{1} gas 
\citep[$\Sigma_{\rm H~I} \approx 1-100\, {\rm M}_\odot \, 
{\rm pc}^{-2}$;][]{jfl+06},
metal-enriched material \citep{sff03}, and large depletion 
factors indicating significant dust formation \citep[e.g.,][]{wfl+06,pcd+07}.
Indirectly, one also infers significant extinction from the reddening of 
some afterglows \citep[e.g.,][]{wfl+06,efh+08}  
and from the large metal column 
densities implied by absorption of the X-ray afterglow 
\citep[e.g.,][]{gw01,crc+06,bk07}.

Evidence for molecular gas, a crucial ingredient of star formation,
has proved elusive thus far.  
\cite{tpc+07} presented a survey for H$_2$ Lyman-Werner
absorption in the afterglow spectra of four GRBs and reported no
positive detections.  \cite{fsl+06} tentatively interpreted an 
absorption feature in their spectrum of GRB~060206 as H$_2$, 
but this may instead be interpreted as an unrelated, \lya\ absorption line.
%Their very low estimates of the molecular fraction
%imply no remaining signature of the molecular cloud that would have beget
%the GRB progenitor; moreover, there are no apparent traces of molecular gas 
%in the gas-rich ISM probed by the sightline.  
The absence of molecules local to the GRB progenitor may be explained
as a natural consequence of photodissociation by the star prior
to its death \citep{wph+08}. 
The absence of molecules along the full sightline, meanwhile, suggests that
an elevated UV radiation field is suppressing
H$_2$ formation throughout the host galaxy
%, an assertion supported
%by UV intensity estimates in GRB hosts 
\citep{cpp+08}.
Nevertheless, these galaxies are actively forming stars \citep{chg04}
and GRB sightlines should occasionally penetrate neighboring 
molecular clouds on the path to Earth.

\begin{figure*}
\begin{center}
\includegraphics[height=6.8in,angle=90]{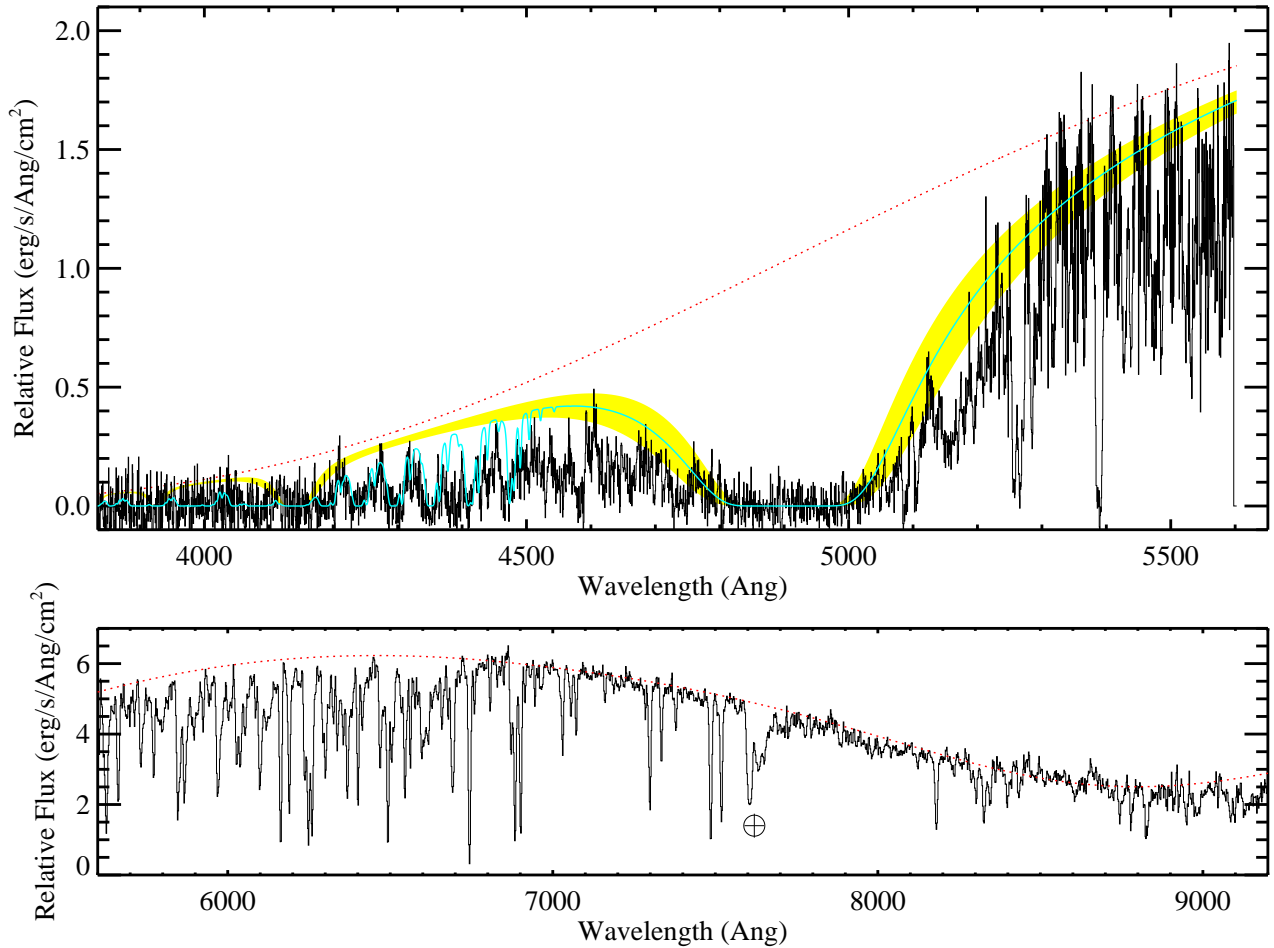}
\end{center}
\caption{Keck/LRIS spectra of the optical afterglow
of GRB~080607 taken with (a) the LRISb camera and the B600
grism and (b) the LRISr camera with the R400 grating.
The red dashed lines indicate our model of
the intrinsic afterglow spectrum reddened by dust in the host
galaxy. 
At $\lambda \approx 4900$~\AA\ one identifies a DLA profile
associated with \ion{H}{1} gas near the GRB.
The shaded region overplotted on the data corresponds to an
\ion{H}{1} column density $\mnhi = 10^{22.7 \pm 0.15} \cm{-2}$.
The model (cyan solid line) includes absorption from H$_2$
Lyman-Werner transitions. 
The line opacity at $\lambda > 5500$~\AA\
is dominated by metal-line transitions from gas in the host galaxy and includes
bandheads of the CO molecule.  Absorption from Earth's the A-band
is marked by a $\oplus$.
}
\label{fig:spec}
\end{figure*}

In this Letter, we present the first such unambiguous detection
of H$_2$ and CO absorption in a GRB host galaxy and provide initial analysis 
of the gas and dust along the sightline to GRB~080607. Previous
detections of intervening CO absorption toward quasars
have only shown trace quantities of
molecular gas \citep{snl+08}.  Emission-line studies of CO and other
molecules, meanwhile, lack sufficient spatial resolution to 
resolve the gas beyond the local universe \citep[e.g.,][]{tnc+06}.
Our data constrain the physical conditions of the ISM including its
metallicity, dust properties, and molecular fractions.
We discuss these results and the prospects for future studies of
molecular clouds using GRB afterglows\footnote{http://www.graasp.org}.

\section{Observations}

We (J.S.B. and D.P.)
initiated spectroscopic observations of the GRB~080607 afterglow with 
Keck~I/LRIS \citep{lris} at 06:27:39 UT on 7 June 2008, $\sim 20$\,min following
the {\it Swift} trigger of the GRB \citep{gcn7847}.  
The Keck/LRIS spectrograph 
has two cameras optimized for blue (LRISb; $\lambda \approx 
3000-5000$~\AA) and red (LRISr; $\lambda \approx 5000-9000$~\AA) optical
wavelengths. Using the $1''$-wide long slit, the atmospheric dispersion corrector,
and the D560 dichroic, we obtained a series of three two-dimensional spectra
of increasing exposure time totaling 2100~s using the B600\,grism in LRISb 
(FWHM $\approx 270~\mkms$)
and the R400 grating in LRISr 
(FWHM~$\approx 300~\mkms$, centered at $\lambda_{\rm center} \approx 7310$~\AA).
We then paused to obtain calibration frames before continuing with
a sequence of two exposures (1500~s each) with the B600 grism and the 
R1200 grating (FWHM~$\approx 110~\mkms$, $\lambda_{\rm center} \approx 6300$~\AA).
All data were obtained %with the slit rotated to the parallactic angle 
under photometric conditions and a seeing better than 1$''$.
The data were reduced and calibrated using the 
LowRedux\footnote{http://www.ucolick.org/$\sim$xavier/LowRedux/index.html .} 
software package developed by S. Burles, J. Hennawi, and J.X.P. using arc images
and a spectrum of HZ44.

\begin{figure*}
\epsscale{1.0}
%\plotone{../Figures/phot_xray.eps}
\plotone{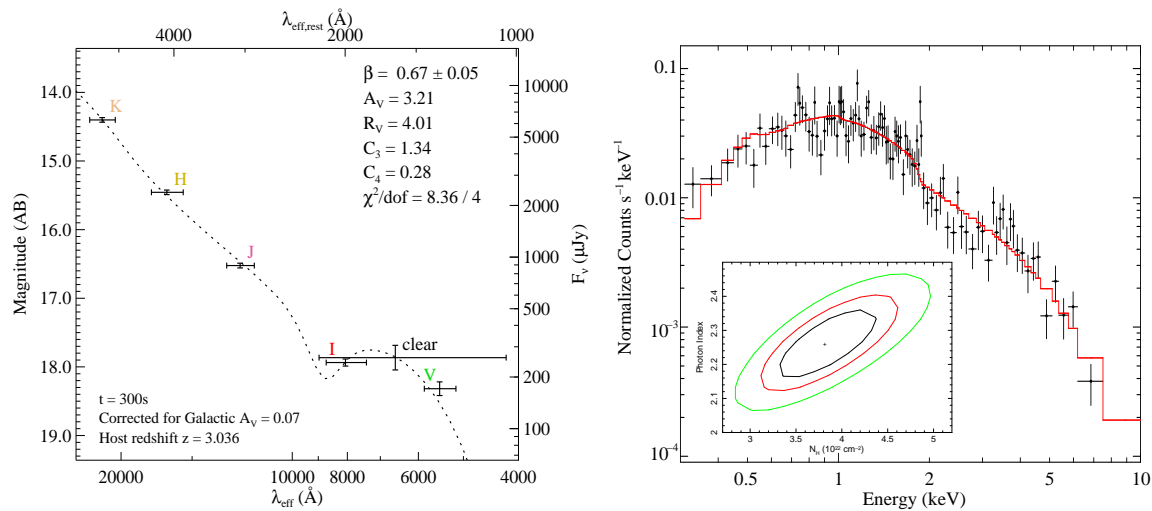}
\caption{(a) Optical/IR photometry for the afterglow of GRB~080607,
as observed by PAIRITEL and KAIT (see \papertwo).
Overplotted on the data is our
best-fit model for an dust-extinguished power-law spectrum, 
$f_\nu \propto \nu^{-\beta}$.  The model, which has a modified
Galactic extinction law, must 
include the 2175~\AA\ feature to match the observations.
(b) {\it Swift}/XRT X-ray spectrum (0.3--10.0 keV) obtained 
0.6--376~ks after the trigger, with the best-fit model overlaid in red. 
The data were modeled as a power law, with a Galactic absorption 
component as well as a redshifted intrinsic absorption component $N_{H}$. 
The best-fit model has an intrinsic 
$N_{\rm H} = 10^{22.58 \pm 0.04} \cm{-2}$ and a photon index 
2.26 $\pm$ 0.07, using the Galactic absorption 
$1.9 \times 10^{20} \cm{-2}$ \citep{dl90}.
The  68\%, 95\%, and 99\% confidence contours of the photon index 
and intrinsic absorption are shown.  
%Overall, the fit gave a C statistic 
%value of 336.3 for 371 degrees of freedom (dof).
}
\label{fig:dustX}
\end{figure*}

The Peters Automated Infrared Imaging Telescope
\citep[PAIRITEL;][]{bsb+06}
responded automatically to the {\it Swift} trigger and 
began simultaneous observations in the $J, H, K_s$ bands at 06:08:44 
UT. A total of 387 individual 7.8~s exposures were taken under moderate 
conditions, seeing $\sim$3$"$, for a total time on target of $\sim$3000~s.
We have reduced these data with the standard PAIRITEL pipeline.
The Katzman Automatic Imaging Telescope \citep[KAIT;][]{fil+01,lfc+03}
also responded to the {\it Swift} alert of GRB 080607 automatically,
and began the first image at 06:09:25 UT, 118 s after the BAT trigger. 
A series of $V$, $I$, and unfiltered images were obtained for the first
$\sim$1800 s after the burst under good conditions.

\section{Analysis}

In Figure~\ref{fig:spec} we present the combined LRIS spectra
of the GRB afterglow. 
A very strong absorption feature is evident
at $\lambda_{\rm obs} \approx 4900$~\AA\ corresponding to damped \lya\
(DLA) absorption from gas within the galaxy hosting GRB~080607.  
The coincidence of strong metal-line transitions including fine-structure 
lines of \ion{Fe}{2} and \ion{Si}{2} confirms this association
\citep{vel+04,pcb06}. 
A Gaussian fit to the strongest low-ion transitions
gives a heliocentric redshift $z_{\rm GRB} = 3.0363 \pm 0.0003$.

The dashed curves represent our model for the 
intrinsic afterglow spectrum, reddened by dust along the sightline.
Assuming an intrinsic optical-infrared
power-law spectral slope $f_\nu \propto \nu^{-\bouv}$, 
we first fit the $JHK$ photometry 
for the single parameter $E(B-V) = 0.8 \pm 0.1$ mag.
%These fluxes probe the rest-frame optical/near-UV, and the analysis
%is nearly independent of the effects of varying 
%far-UV extinction laws.  
In this analysis we 
accounted for the uncertainty in the unknown intrinsic 
spectral index, which for 
afterglows at early times is found to be in the range $\bouv = 0.1-1.1$
\citep{kkz+07}. 
We then fitted a \cite{fm90} dust model to 
unabsorbed regions of the R400 spectrum (correcting for Galactic extinction), 
holding $E(B-V)$ fixed but allowing all other parameters to vary\footnote{
This allows for color evolution between the times of the photometric
and spectroscopic observations.}.
For an assumed $\bouv = 0.5$, our best-fit parameters are 
$R_V = 4.0 \pm 0.23$, $A_V = 3.2 \pm 0.5$\,mag, $c_3 = 1.3 \pm 0.3$, 
and $c_4 = 0.3 \pm 0.1$ (Fig.~~\ref{fig:dustX}a).
%where the last two values parameterize (Figure~\ref{fig:dustX}a) 
%the 2175~\AA\ bump strength and the UV extinction, respectively.

Although there is considerable degeneracy among
these values, we reach three robust conclusions:
(i) the $R_V$ value is higher than the typical Galactic and SMC values;
(ii) the data require a nonzero $c_3$ parameter indicating the positive
detection of the 2175~\AA\ bump, only the second detection of this feature
associated with a GRB host galaxy \citep{kkg+08,efh+08}; and
(iii) the data favor a low $c_4$ value suggesting few small dust grains.
These characteristics are distinguished from the extinction laws commonly
inferred for GRB sightlines which generally have SMC-like or ``grey'' 
extinction curves and lack evidence for a 2175~\AA\ bump 
\citep[e.g.,][]{kkz06,bph+06,wfl+06,pbb+08}.
%The photometric data
%and the extinguished afterglow model are presented in
%Fig.~\ref{fig:dustX}a.  This same model is overplotted on the
%spectral data in Fig.~\ref{fig:spec}, where we assumed negligible
%line opacity at $\lambda \approx 5330$~\AA\ and 7140~\AA\ to 
%normalize the model.

With the reddened afterglow spectrum as our input continuum, 
we have fitted the \lya\
absorption to the data redward of 5000~\AA\ with centroid 
$\lambda = 1215.67(1+z_{\rm GRB})$~\AA. 
The best-fit value for the \ion{H}{1} column density is
$\mnhi = 10^{22.70 \pm 0.15} \cm{-2}$, where the {\it uncertainty} is dominated
by the continuum model. 
%Blueward of the DLA profile, the flux is severely reddened
%by dust extinction.
%Nevertheless, 
We detect the afterglow down to at least $\lambda \approx 4000$~\AA, primarily
in a series of emission ``gaps'' 
(e.g., at $\lambda_{\rm obs} \approx 4200, ~4275, ~4315$~\AA).
Although the absorption between the gaps
could be strong IGM absorption, this would require 
a much higher integrated opacity than observed along quasar 
sightlines. We interpret the majority of this opacity as 
Lyman-Werner bands of molecular hydrogen at $z=z_{\rm GRB}$.
	
\begin{deluxetable}{lcccc}
\tablecaption{EW Measurements\label{tab:ew}}
\tablewidth{0pt}
\tablehead{\colhead{Region [\AA]} & 
\colhead{$\lambda_{\rm c}^a$ [\AA]} & 
\colhead{W$_{\rm obs}$ [\AA]} & \colhead{ID} & \colhead{$z_{abs}$}}
\startdata
\cutinhead{B600}
$\lbrack$5045.3, 5053.5] &&$ 6.73\pm0.27$&SII 1250&3.037\\
$\lbrack$5055.7, 5065.0] &&$ 7.58\pm0.28$&CrII 2056&1.462\\
&&&SII 1253&3.037\\
$\lbrack$5081.2, 5093.6] &&$10.66\pm0.33$&CrII 2066&1.462\\
&&&SII 1259&3.037\\
\enddata
\tablenotetext{a}{Central wavelength determined from a Gaussian fit.  $W_{\rm obs}$ values determined from boxcar integrations 
have no value listed.}
\tablecomments{[The complete version of this table is in the electronic edition of the Journal.  The printed edition contains only a sample.]}
\end{deluxetable}

\begin{figure}
\begin{center}
\includegraphics[width=3.5in]{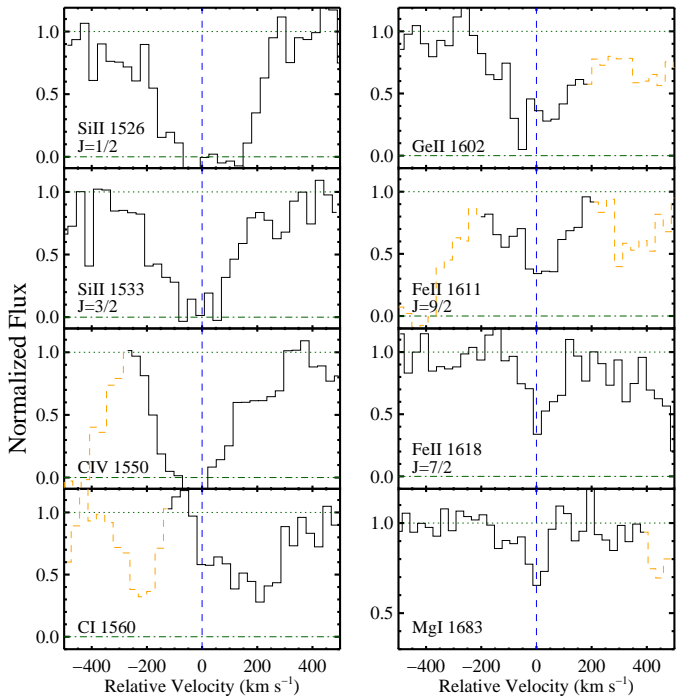}
\end{center}
\caption{Velocity plots of a subset of the atomic and ionic transitions
associated with the ISM of the galaxy hosting GRB~080607.  
In all panels, $v=0 \mkms$ corresponds to $z_{\rm GRB} = 3.0363$ and the
orange dashed lines indicate blends between transitions.
}
\label{fig:metals}
\end{figure}

The LRISr/R400 spectrum (Fig.~\ref{fig:spec}b)
covers rest wavelengths
$\lambda_{\rm rest} > 1300$~\AA\ relative to the GRB and, as such, 
{\it none of the absorption} is related to the \lya\ forest despite
its similar appearance.
Instead, the features are metal-line
absorption by gas in the GRB host galaxy (Fig.~\ref{fig:metals})
and/or intervening gas.
We have measured equivalent widths ($W_\lambda$) for these features
by fitting Gaussian profiles or, in cases of severe line-blending,
by boxcar summation over the spectral window (Table~\ref{tab:ew}).
The absorption is dominated by 
resonance transitions and fine-structure lines of abundant atoms and ions
at $z=z_{\rm GRB}$, but we also identify absorption from two strong 
\ion{Mg}{2} absorbers ($W_{2796} > 1$~\AA)
at $z = 1.340$ and 1.462, an unusually common   
characteristic of GRB afterglow spectroscopy \citep{ppc+06}.

The GRB absorption features are highly saturated and one
can only estimate lower limits to the ionic column densities.  
Under the {\it very} conservative assumption that 
$N = 10^{23.05} W_\lambda/(f \lambda^2) \cm{-2}$ 
(i.e., the weak limit of the curve of growth, with $f$ 
the oscillator strength),
we infer [S/H]~$> -1.7$, [Si/H]~$> -2$, [Fe/H]~$> -1.8$, and [Zn/H]~$>-1.7$.  
%Integrating the apparent optical depth profiles \citep[][]{savage91},
%we derive lower limits that are 0.3 to 0.5\,dex larger.
We also detect very weak transitions of several common elements
(e.g., \ion{O}{1}~1355, \ion{Ni}{2}~1804, \ion{Mg}{1}~1683, \ion{Co}{2}~1574)
and a few transitions from rare elements (e.g., \ion{Ge}{2}~1602).
The equivalent width of the \ion{Ge}{2}~1602 transition alone
implies a lower limit [Ge/H]~$> 0$ and analysis of \ion{Co}{2} transitions,
a refractory element, yields [Co/H]~$>-0.7$\,dex.   
This is consistent with the equivalent
width of the absorption coinciding with the \ion{O}{1}~1355 transition
which implies [O/H]~$> -0.2$\,dex.
We conclude that the integrated gas metallicity is solar
or even super-solar, a result that challenges the prevailing
collapsar model \citep[e.g.,][]{wh06}. 

\begin{figure}
\begin{center}
\includegraphics[width=3.5in]{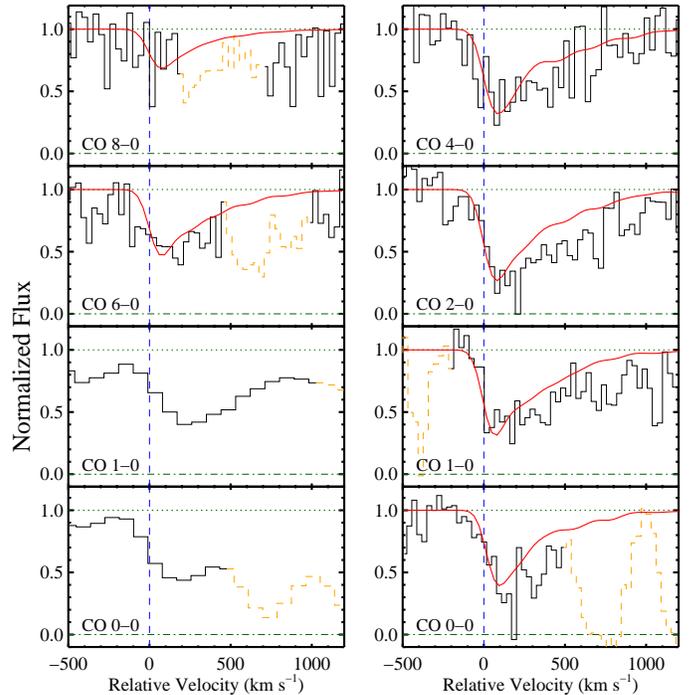}
\end{center}
\caption{Velocity plots of the molecular transitions
associated with the ISM of the galaxy hosting GRB~080607.  
The first column shows the CO $A-X$ bandheads taken with the LRISb/B600
grism (top two) and LRISr/R400 grating (bottom two).
All other data refers to the LRISr/R1200 spectrum.
The red line overplotted on the CO data indicate our best-fit
model corresponding to
$\N{\rm CO} = 10^{16.5} \cm{-2}$, $b_{\rm CO} = 2.0 \mkms$, 
$\texco  = 300$~K, and a velocity offset from $z_{\rm GRB}$ of
$\delta v_{\rm CO} = 30$~km s$^{-1}$.
%We interpret the mismatch between the data and model for
%the CO 2-0 bandhead as coincident absorption from an unknown
%source of opacity.
In all panels, $v=0 \mkms$ corresponds to $z_{\rm GRB} = 3.0363$ and the
orange dashed lines indicate blends between transitions.
}
\label{fig:CO}
\end{figure}

The optical spectra also reveal the 
positive detection of a series of CO $A-X$ bandheads (Fig.~\ref{fig:CO}).
%In Fig.~\ref{fig:velplt} we present a velocity plot of these bandheads
%together with a subset of the atomic and ionic transitions.  
It is noteworthy that the peak depth of the bandheads is offset by
$\delta v \approx + 150\mkms$ from the atomic transitions and 
that the bandheads show absorption to $\delta v \approx +1000 \mkms$.
At first glance, this suggests that the molecular gas 
has an extremely high velocity dispersion,  
%Although we cannot strictly eliminate
%such a model with data at our spectral resolution,  
a highly unlikely attribute for molecular gas.  
We interpret these unusual
characteristics, instead, as the result of absorption from high
rotational levels ($J>10$) which have been populated by UV pumping
from the GRB afterglow.  This is analogous to the excitation of
fine-structure levels in Fe$^+$, Si$^+$, and O$^0$ gas in other
GRB galaxies \citep{pcb06,vls+07}. Such levels would be easily
resolved with an echelle spectrum.

We have fitted the CO $A-X$ bandheads in the R1200 and B600 data
with a series of models and find preferred values
$\N{\rm CO} = 10^{16.5 \pm 0.3} \cm{-2}$, $b_{\rm CO} = 2.0 \pm 0.3 \mkms$, 
$\texco  = 300_{-70}^{+200}$~K, and a velocity offset from $z_{\rm GRB}$ of
$\delta v_{\rm CO} = 30 \pm 15 \mkms$ (Fig.~\ref{fig:CO}).  
The preferred values were found by minimizing the residuals between
the data and the model for the 1-0, 2-0, and 4-0 CO bands 
simultaneously. Each model included
all rotational levels up to $J = 25$, for a total of 225 CO transitions with
transition data taken from fits to empirical data \citep{mn94}. 
We then visually compared the model prediction 
with the 6-0 and 8-0 CO bands seen in the B600 spectrum
to assure consistency over a larger range of $f$-values;  
these bandheads set an upper limit of $\N{\rm CO} < 10^{17.2} \cm{-2}$.
Although a parameterization of the gas with an excitation
temperature ($\texco$) is not formally valid for a photon-pumped gas,
we achieve a reasonable model.
%This value of $\texco$ is higher by a factor of $\sim$30 than
%that expected for photoexcitation induced by the cosmic microwave
%background radiation at $z_{\rm GRB}$ 
%\citep[$T_{\rm CMB} = 11.0$~K; 
%see also][]{snl+08}.

%We have not included the R400 data in our analysis, in part because
%we observe small but statistically significant differences between the
%normalized profiles (especially the CO 4-0 bandhead).   These differences
%may be attributed to uncertainties in the continuum, but it is also
%possible that we are observing temporal variations in the populations
%of the rotational levels of the ground state.   Observations
%with greater spectral resolution are necessary to determine the underlying
%cause of these variations. 

We have analyzed the H$_2$ gas in a similar fashion.  Because the
Lyman-Werner lines are heavily saturated and blended, there is little
sensitivity to the Doppler parameter $b_{\rm H_2}$.  Adopting
$b_{\rm H_2} = 3\mkms$, we varied $\texh$ and \nht\ and compared
these models against the data at $\lambda_{\rm rest} < 1100$~\AA.
Under the assumption that H$_2$ absorption dominates the opacity
at $\lambda_{\rm obs} < 4200$~\AA,
we constrain $\mnht = 10^{21.2 \pm 0.2} \cm{-2}$ 
and $\texh = 10$ to 300~K.  Models with 
larger/smaller \nht\ values conflict with the observations.
%$\mnht > 10^{21.7} \cm{-2}$ substantially overpredict absorption at
%$\lambda_{\rm obs} = 4213, 4262$~\AA\ where positive flux is detected.
%Models with $\mnht < 10^{20.7} \cm{-2}$ would imply detectable flux
%at $\lambda_{\rm obs} < 4200$~\AA.  
Our analysis sets an upper limit to
the integrated molecular fraction $f_{\rm int}({\rm H_2}) < 0.13$, but we
expect that the sightline is comprised of at least two clouds:
one predominantly atomic and the other molecular.

Assuming solar relative abundances,
the implied O column density is $\N{O} > 10^{19.6} \cm{-2}$
for a Galactic equivalent column density $N_H > 10^{22.7} \cm{-2}$.
This metal column density implies that there should be
significant absorption of 
soft X-rays by K-shells of O, Si, etc.  In Fig.~\ref{fig:dustX}b,
we present the X-ray spectrum and best-fit model of the {\it Swift}/XRT
data obtained 0.6--376~ks after the burst \citep{gcn7955}.
We estimate an effective $N_{\rm H} = 10^{22.58 \pm 0.04} \cm{-2}$
and a photon index $\Gamma = 2.26 \pm 0.07$ for a power-law spectrum.
Therefore, the metal column
densities from analysis of both the UV and X-ray spectra are consistent.
On the other hand, assuming the Galactic dust-to-metals ratio would give
an extinction ($A_V > 10$~mag) inconsistent with the observations.  
We conclude that the gas
has a much lower dust-to-metals ratio than Galactic
\citep{gw01,wfl+06}.
%It is unlikely that this is due to the preferential destruction of dust
%by the GRB afterglow \citep[e.g.,][]{fkr01,draine02} because, as argued
%in the following section, the majority of gas lies too far ($r > 100$\,pc)
%from the afterglow. 
%Instead, we speculate that the majority of dust extinction is associated
%with a lower column density molecular cloud distinct from atomic
%gas that is nearly dust-free.

%Finally, we note many lines that remain unidentified in Table~\ref{tab:ew}.
%It is our expectation that the majority of these are atomic transitions
%of common elements (e.g., \ion{Ni}{1}, \ion{Si}{1}, \ion{Fe}{1}) and
%weak transitions of several low-ionization species (e.g., \ion{Fe}{2}, 
%\ion{Ni}{2}).

\section{Discussion}

The previous section described the physical properties of the ISM
and the first clear detection of molecular gas in a GRB host galaxy.
With an $A_V \approx 3.2$~mag, this represents a molecule-rich
``translucent'' sightline with effective UV shielding of CO \citep{vb88}.
With $\N{\rm CO} \approx 10^{16.5} \cm{-2}$ and 
$\mnht  \approx 10^{21.2} \cm{-2}$, the abundance
ratio of CO to H$_2$ is $\sim 2 \times 10^{-5}$. These values are consistent
with detailed models of CO photodissociation in translucent environments
\citep[e.g., model T3 in][]{vb88},
and with the transition region between diffuse and dark clouds
in the correlation plot of CO versus H$_2$ \citep{srf+08}.
%Remarkably, these results are similar to translucent environments
%within the Milky Way.
Altogether, the gas exhibits roughly solar metallicity, the dust has 
an extinction curve similar to that observed for Galactic molecular clouds,
including a 2175~\AA\ bump, and the CO/H$_2$ ratio is consistent
with Galactic translucent clouds.
We infer that at least some molecular clouds in the young universe 
have similar properties to those observed locally.  

There are, however, two notable
differences from the Galactic ISM.  First,  we observe a very large 
\ion{H}{1} column density.  This attribute is common to GRBs 
yet contrasts with the surface densities of local \ion{H}{1} disks.
While the formation of a molecular cloud requires a large dust
column to absorb H$_2$-dissociating photons,
the analysis of \cite{kmt08} implies a 20 times
lower column than observed. 
Perhaps the large \nhi\ value is a coincidence;
these are routinely observed along GRB sightlines without molecules
\citep{tpc+07}
and  our preferred model of the CO absorption suggests a velocity
offset of $\delta v_{\rm CO} \approx +30\mkms$ from the atomic metal lines.
The second key difference is that we infer a much lower
dust-to-metals ratio than Galactic whether that metal column is
estimated from the UV or X-ray spectra.
We interpret this difference as the result of a dust-poor, atomic
phase that is spatially separated from the highly extinguished molecular cloud.
Both phases of gas give rise to metal absorption (the profiles
span the estimated $30 \mkms$ offset; Fig.~\ref{fig:metals})
but we speculate that the atomic phase has a much lower dust-to-gas ratio.

% THIS COULD BE SHORTETEND A LOT
Several lines of evidence argue against identifying 
this molecular cloud as the birthplace of the GRB progenitor.
We detect strong absorption from \ion{C}{1} and
\ion{Mg}{1} transitions indicating  
%While C$^0$ atoms are somewhat shielded
%by dust and H$_2$ line opacity, we detected copious
%photons at $h \nu \gtrsim 8$\,eV from the afterglow. 
the gas lies $>100$\,pc from the GRB \citep{pcb06}.  If the molecular
gas were within $\sim 10$\,pc of the afterglow, a significant
column of H$_2$ would be excited to unbound states \citep{draine02}.
We have searched for such transitions at $\lambda_{\rm rest} > 1400$~\AA\
and report no positive detections, setting an equivalent width limit
$W_{\rm rest}({\rm H_2^*}) < 0.1$~\AA\ (95$\%$ c.l.).
We have searched for variability in the dust opacity,
\ion{H}{1} absorption, and in the equivalent widths of the metal-line
transitions but detect none.  
%This further indicates the gas lies well beyond the GRB afterglow.
The most plausible scenario, therefore, is that the 
gas lies beyond
100\,pc from the afterglow, consistent with the estimate for 
photodissociation regions surrounding GRB progenitors \citep{wph+08}.
%We can place an upper limit on the distance to the cloud based on the
%excitation of Fe$^+$ \citep{pcb06} and the absorption by upper $J$ levels
%of CO.  We estimate a distance of less than a few kpc, provided we adopt
%the observed (extinguished) flux from the afterglow.
We conclude that the molecular cloud identified in the afterglow
spectrum of GRB~080607 arises in a neighboring, star-forming region.

This marks the first unambiguous detection of 
H$_2$ in a GRB sightline (among six intensively searched) and of CO
(among $\sim 20$ sightlines).  Our results offer insight
into the rarity of these detections.  From our dust model, we
estimate a rest-frame UV extinction of 
$A_{1100 \rm \AA} \approx 8$\,mag.
If this afterglow were just 1~mag fainter, then
the Lyman-Werner series could not have been detected.
Similarly, the CO bandheads would not have been observable
for the majority of GRBs.
We conclude, therefore, that one requires an uncommonly
luminous afterglow combined with
a fortuitous path through an unrelated molecular cloud.
%This may help to explain the near absence of any such detections along the
%thousands of quasar sightlines to date \citep{zp06,nlp+08}.

We contend further that a significant fraction of 
the subluminous optical bursts 
\citep[also called ``dark bursts'';][]{dfk+01,jbf+05,ckh+08}
are events like GRB~080607 but with 
fainter intrinsic luminosity.  An open and intriguing question
is whether these dust-extinguished events correspond to unique host galaxies
(e.g., intensely star forming and dusty systems),
or whether the hosts have properties similar to the general population 
and the variations along individual sightlines
is geometrical (i.e., molecular clouds have small cross-section
and are rarely intersected).  A resolution of these two effects
bears on the nature of molecular gas at high redshift including its
distribution and relation to star formation.
A multi-wavelength survey of the hosts, including searches
for CO emission, may help distinguish between these scenarios.

Though rare, these high-$z$ events enable studies at optical wavebands
of rest-frame UV features that serve as powerful diagnostics
of star-forming regions.  This includes a precise redshift for follow-up
observations, for example studies of CO in emission at millimeter wavelengths. 
Indeed, the CO 3-2 and 1-0 transitions lie at favorable observational frequencies.
High-resolution spectroscopy of systems like GRB~080607 would also
permit searches for additional molecules in the UV \citep[e.g., C$_2$;][]{swt+07}.
%We have performed a search in our spectrum without success.
Finally, such systems easily satisfy the metal-strong DLA criteria
of \cite{shf06} and afford a special opportunity to study
rare elements in the young universe.  Assuming solar abundances,
for example, 
we estimate that the \ion{Sn}{2}~$\lambda$1400, \ion{B}{2}~$\lambda$1362, 
and \ion{Pb}{2}~$\lambda$1433 transitions should all have rest-frame 
equivalent widths exceeding 50~m\AA.
%We anticipate future observations at higher spectral
%resolution than those presented here for GRB~080607.

\acknowledgements 
J. X. P. is partially supported by NASA/{\it Swift} grant
NNX07AE94G, and an NSF CAREER grant (AST--0548180).
A.V.F. is grateful for NSF grant AST--0607485 and the TABASGO Foundation.
Most of the data presented herein were obtained at the W. M. Keck
Observatory, which is operated as a scientific partnership among the
California Institute of Technology, the University of California, and
the National Aeronautics and Space Administration (NASA). 
%The
%Observatory was made possible by the generous financial support of the
%W. M. Keck Foundation. 
KAIT has been supported primarily by donations from AutoScope Corp., 
Lick Observatory, the NSF, 
the Sylvia \& Jim Katzman Foundation, and the TABASGO Foundation. 
%ongoing operation were made possible by donations from Sun
%Microsystems, Inc., the Hewlett-Packard Company, AutoScope
%Corporation, Lick Observatory, the NSF, the University of California,
%the Sylvia \& Jim Katzman Foundation, and the TABASGO Foundation. 
The PAIRITEL project is operated by the
Smithsonian Astrophysical Observatory (SAO) and was made possible
by a grant from the Harvard University Milton fund, the camera loan
from the University of Virginia, and the continued support of the SAO
and UC Berkeley; it is further supported by NASA/{\it Swift}
Guest Investigator Grant \#NNG06GH50G.  
We thank M. Skrutskie and A. Szentgyorgyi for their support of
PAIRITEL.
We are grateful to E. E. Falco
and the Mt. Hopkins staff (W. Peters and T. Groner) for their continued
assistance with PAIRITEL.
This work made use of data supplied by the UK Swift Science Data Centre at 
the University of Leicester.
We acknowledge D. Welty and M. Krumholz for valuable discussions.

%\bibliographystyle{/u/xavier/paper/Bibli/apj}
%\bibliography{/u/xavier/paper/Bibli/allrefs}

\end{document}